\documentclass[10pt]{article}

\usepackage{amsmath}
\usepackage{amssymb}

\usepackage{graphicx}
\usepackage{cite}

\usepackage{color}

\usepackage[labelfont=bf,labelsep=period,justification=raggedright]{caption}

\makeatletter
\renewcommand{\@biblabel}[1]{\quad#1.}
\makeatother
\date{}

\newcommand{\n}{seven}
\newcommand{\N}{Seven}
\newcommand{\sa}{\emph{Staphylococcus aureus}}
\newcommand{\website}{http://opencfu.sourceforge.net}
\newcommand{\humantime}{39.3 (sd=6.1)}
\newcommand{\IJM}{IJM}

\newcommand{\EdgeNoiseOCFU}{$0.159$ ($sd=1.64, P$-$value=0.34$)}
\newcommand{\EdgeNoiseIJM}{$26.0$ ($sd=19.7, P$-$value=9.01\cdot 10^{-6}$)}
\newcommand{\EdgeNoiseNICE}{$7.37$ ($sd=7.38,P$-$value=1.9\cdot 10^{-4}$)}

\newcommand{\BubbleNoiseOCFU}{$R^2 = 0.00$ ($a=0.03, P$-$value=0.90)$ }
\newcommand{\BubbleNoiseNICE}{$R^2 = 0.65$ ($a=0.94, P$-$value=3.35\cdot 10^{-5}$) }
\newcommand{\BubbleNoiseIJM}{$R^2 = 0.66$ ($a=1.18, P$-$value=2.56\cdot 10^{-5}$) }

\begin{document}

\begin{flushleft}
{\Large
\textbf{OpenCFU, a New Free and Open-Source Software to Count Cell Colonies and
Other Circular Objects}}

Quentin Geissmann$^{1,2,\ast}$
\\

\bf{1} Department of Animal and Plant Sciences, University of Sheffield,
Sheffield, United Kingdom, \bf{2} Institute for Biology, Free University of 
Berlin, Berlin, Germany. \\ $\ast$ E-mail: Corresponding
q.geissmann@fu-berlin.de
\end{flushleft}

\section*{Abstract}

Counting circular objects such as cell colonies is an important source of
information for biologists.
Although this task is often time-consuming and subjective, it is still
predominantly performed manually. The aim of the present work is to provide a
new tool to enumerate circular objects from digital pictures and video streams.
Here, I demonstrate that the created program, OpenCFU, is very robust, accurate and
fast. In addition, it provides control over the processing parameters and is
implemented in an intuitive and modern interface. OpenCFU is a cross-platform
and open-source software freely available at \website{}.

\section*{Introduction}

Counting objects has always formed an important element of data collection in
many fields of biology.
It is therefore very common for
biologists to enumerate objects such as pollen\cite{costa_counting_2009},
eggs\cite{mello_image_2008}, seeds\cite{severini_counting_2011},
nuclei\cite{forero_deadeasy_2010}, cells\cite{kachouie_arraycount_2009} or organisms\cite{yati_flycounter:_2011}.
Given that
such tasks are time-consuming and, to some extent, subjective, it is
surprising that automation is still infrequent. 
Effectively, enumerating objects is a two-part process: image capture and image
analysis. Nowadays, technologies such as digital cameras and webcams provide
an increasingly high image quality and are increasingly inexpensive.
Simultaneously, many optimised image processing
algorithms and open-source libraries can be used on laptops
and desktop computers.

In different fields of microbiology, immunology and cellular biology, counting
colonies of cells growing on agar plates is routine.
Automating such counting procedures is not simple since colonies must first be isolated
from the background and then, if they overlap, be separated. In addition, such methods must be capable of rejecting common artefacts
such as imperfections in the agar, dust and edges of Petri dishes.
However, since cell colonies are topologically fairly simple objects, solutions to
enumerate them from pictures have long since been considered
\cite{mansberg_automatic_1957,mukherjee_bacterial_1995}. Commercial tools have
been developed\cite{putman_simplified_2005}, but remain expensive.
Furthermore, the fact that the programs they provide are proprietary 
(\emph{i.e.} not open-source) makes them very restrictive: it is rarely possible
to know the precise nature of the analysis being performed, nor to modify or
share these programs.
In contrast, some authors have recently published open-source methods to count
colonies from digital images
\cite{cai_optimized_2011,bewes_automated_2008,niyazi_counting_2007,clarke_lowcost_2010,sieuwerts_simple_2008,vokes_using_2008,brugger_automated_2012}. 
These use a variety of techniques such as generalised Hough transform\cite{bewes_automated_2008}, 
template matching algorithm \cite{kachouie_arraycount_2009},
advanced illumination correction and particle filter \cite {vokes_using_2008},
watershed algorithm on a thresholded image \cite{sieuwerts_simple_2008,cai_optimized_2011} and,
more recently, a two-pass thresholding procedure in combination with statistical models \cite{brugger_automated_2012}.
They were either implemented as
standalone programs capable of analysing image files \cite{niyazi_counting_2007,bewes_automated_2008,kachouie_arraycount_2009,clarke_lowcost_2010},
extensions to existing image processing programs \cite{vokes_using_2008,sieuwerts_simple_2008,cai_optimized_2011} or
integrated (hardware-software) counting systems \cite{brugger_automated_2012}.

All these methods claim to give results very comparable to human counts, to improve
objectivity and save time. However, none of them have been widely adopted.
In addition to the fact that some biologists are not aware of the existence of such
tools, their apparent failure to be adopted could be explained by shortcomings
in performance, such as an inability to split merged colonies
or long processing times. It is also very important for a method to be robust and versatile:
it must be able to perform well without changing parameters when optimal conditions vary slightly,
and the presence of artefacts should only have minor consequences.
Another reason for non-adoption of automatic methods could be their lack of
user-friendliness: even a perfect method may fail to be adopted if the software
that implements it does not provide a modern user interface. Such an interface
should enable the user to easily submit a list of images for analysis and to
manually alter the values of the processing parameters. Additionally, the user
should be able to select a region of interest and visually check the results.
Since many users do not have deep knowledge of the underlying image processing, 
it is also important that the parameters are intuitive and few.
A final reason could be the lack of public availability
and maintenance. It is not rare that authors publish work about a
program without mentioning a download link, releasing the software on a public
repository or enclosing it as a supplementary
material\cite{bewes_automated_2008,kachouie_arraycount_2009}.
In addition, a program will most likely need to be updated to correct
unpredicted behaviour or avoid reliance on obsolete dependencies.

The aim of the work presented herein is to provide an alternative open-source tool that
features very robust, accurate and fast image processing as well as a modern and
functional user interface. Thanks to the optimised OpenCV
library\cite{opencv_library}, a rapid implementation of image processing
functions has been possible. This renders the analysis of very large and
numerous picture-sets easy and offers integration of capture devices 
(such as webcams). 
In order to assess the relative efficacy and usefulness of the new software,
a comparison to two other available tools was undertaken.

The NIST's Integrated Colony Enumerator (NICE)\cite{clarke_lowcost_2010}
software
implements a combination of threshold and extended minima in order to improve robustness.
The program comes with a functional user interface relying on few intuitive parameters
and provides accurate results. 
However, it is relatively slow and lacks
postprocessing filters.
More recently, Cai \emph{et al.} published\cite{cai_optimized_2011} a short
ImageJ macro which they described as very accurate and only depending on two parameters.
The method relies on adaptive thresholding, watershed algorithm and subsequent particle filtering.
It does not provide a functional user interface or indicate a way to open a list
of files and display the outlines of the detected colonies, nor does it allow
postprocessing of the detected colonies. 
Since ImageJ macro language is
flexible, it is possible, but requires some familiarity with the language.

My results show that OpenCFU, the software created, is faster, more accurate and
more robust to the presence of usual artefacts than NICE and Cai \emph{et al.}'s macro.
As well as efficiently counting
bacterial colonies, the program can also 
be used to enumerate
other circular objects such as seeds or pollen. 

\section*{Algorithm and Implementation}

\subsection*{Algorithm} 
\begin{figure}[!ht]
\begin{center}
\includegraphics[width=4.27in]{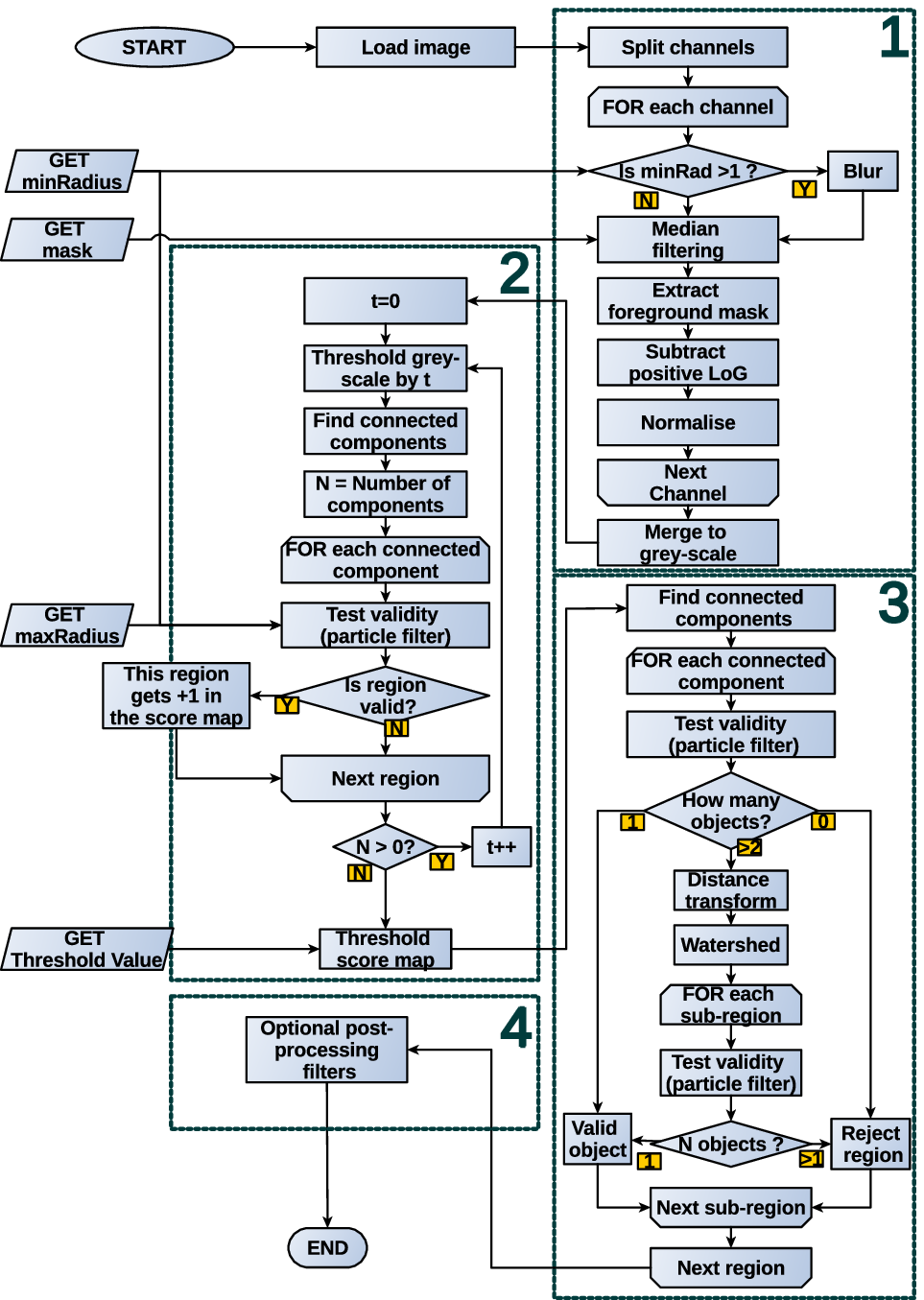}
\end{center}
\caption{{\bf Flowchart representing the processing steps.}
The image is preprocessed (1) in order to correct for gradual
changes in background intensity and increase the contrast.
The first pass of the processing (2) generates a score-map by iteratively annotating valid regions.
The second pass (3) involves finding connected components in the thresholded score-map and segmenting them using a distance transform/watershed approach.
Optional postprocessing filters (4) can be performed by OpenCFU or, using the raw data, by the user. 
}
\label{figFlowChart}
\end{figure}

\begin{figure}[!ht]
\begin{center}
 \includegraphics[width=3.27in]{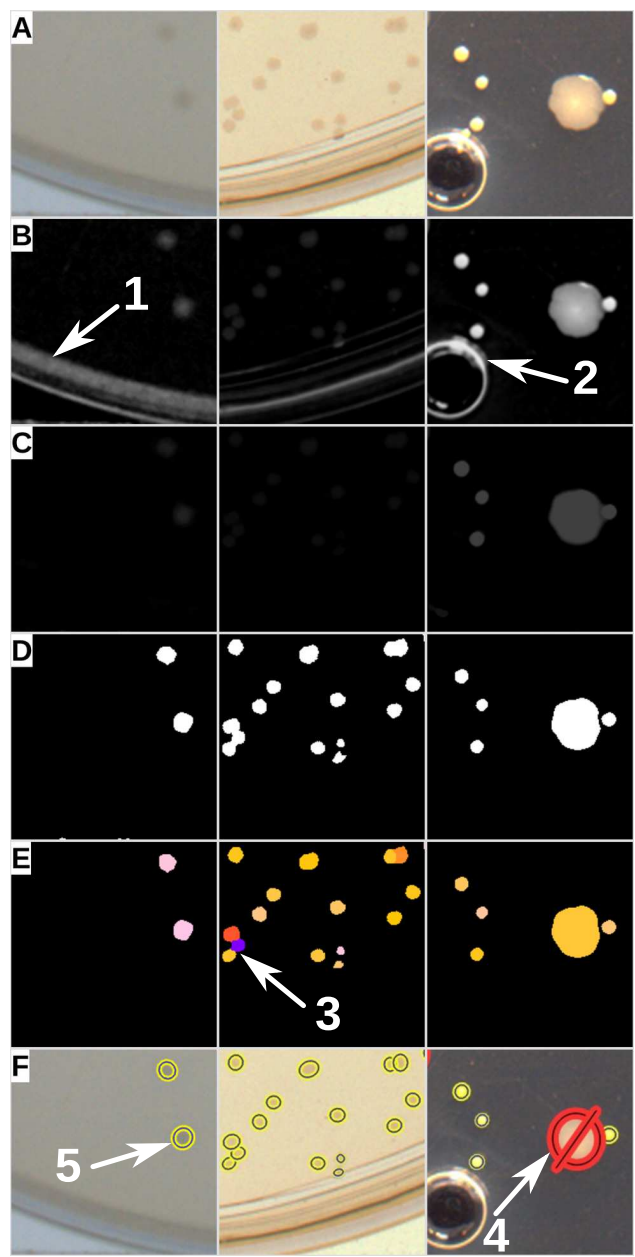}
\end{center}
\caption{{\bf Illustration of the processing steps performed on three sample images.}
Each channel of the original image (A) is preprocessed individually and merged
to a grey-scale image~(B).
A score-map is generated by recursive thresholding an annotation of circular regions~(C). 
This excludes regions that were morphologically unlikely to be colonies~
(\emph{i.e.} arrows 1 and 2).
The score-map is then thresholded by a user-defined or automatically calculated
value~(D).
The objects identified as merged colonies (on the basis of their morphological
features) are segmented using a watershed variant on their distance transform~(E).
Arrow 3 shows objects that have been successfully segmented. Finally, the morphologically valid objects can be assessed further using intensity and colour filters.
Arrow 4 shows a minority contaminant bacteria that was excluded using postprocessing filter and represented by crossed-out red ellipses. 
Arrow 5 shows valid colonies represented by yellow and blue ellipses.
For the purpose of explanation, only representative areas (200$\times{}$200 pixels)
of three processed images are shown here.}
\label{figPipeline}
\end{figure}

Methods relying on direct thresholding of a grey-scale image followed
by morphological segmentation are likely to detect high contrast artefacts such
as parts of the edges of Petri dishes and bubbles.  For a given value of threshold,
some portions of these artefacts could, by chance, be morphologically similar to colonies and will result in
detection of false positives.
The algorithm proposed here aims to increase robustness by virtually testing
all possible values of threshold and keeping only regions that were recurrently morphologically
valid. A detailed flowchart is provided in figure~\ref{figFlowChart} and portions of images 
at different processing stages are shown in figure~\ref{figPipeline}.
Briefly, the colour image is split into three channels from which the background
is independently estimated using a local median filter.
In order to enhance discrimination of foreground objects,  the positive
Laplacian of Gaussian of each channel's foreground mask is subtracted from
itself and three preprocessed channels are normalised and merged to form a
new grey-scale image.

The first pass of the processing involves thresholding the resulting grey-scale
image by multiple values.
For each value, the algorithm will search for connected components in the binary
image.
Each component is assessed by a particle filter that takes into account relationships
between variables such as area, perimeter, convexity, aspect ratio and
hollowness in order to determine whether or not a region is likely to be valid 
(\emph{i.e} made from one or more circular objects).
Over the range of threshold values, every time a valid region is found, all
its pixels are incremented in a ``score-map''.
The score-map can be understood as a representation of how recurrently (over the
iterations of threshold) pixels are part of a circular region. Finally,  a
user-defined (or automatic) threshold is applied to the score-map.
During the second pass of the processing, a similar particle filter is applied
but this time it classifies the connected components as ``invalid'',
``individual object'' or ``multiple objects''.
``Individual objects'' are accepted and ``invalid'' ones  are rejected
immediately, whilst ``multiple objects'' are morphologically segmented using a variant
of the watershed algorithm on their distance-map\cite{marotz_effective_2001}.
Finally, all segmented objects are reassessed by the particle filter.

Optionally, a normal distribution is fitted to the relative colour intensities of
objects that were not split (since they are less often falsely positive). This
distribution then serves to compute the likelihood of each object to be valid.
Finally, a simple likelihood user-controlled threshold is applied to exclude
marginal objects.

 \subsection*{Implementation and User Interface} For performance reasons, OpenCFU was
programmed in \texttt{C++}. The image processing was implemented using OpenCV framework\cite{opencv_library}
 which offers highly optimised image processing functions.
The time-consuming loops were optimised further for multi-core architecture using
OpenMP library\cite{openmp11}. The graphical user interface was designed using GTKmm.
These three libraries are open-source, cross platform and regularly maintained.

OpenCFU was designed in order to accelerate the calibration phase by having a
fast processing time and by immediately displaying results after parameters have
been changed. In addition, when a parameter is changed, OpenCFU dynamically restarts the analysis from
the first step involving this variable, as opposed to systematically re-analysing the image
from 
the beginning. For instance, if the value of a postprocessing filter is altered, the whole processing will not be re-run.
The software also integrates an interface to video
devices and
other features such as optional
automatic dish detection based on Hough circle transform.

The program can deliver two different types of result: a summary or a detailed 
output. In the summary, each row of data contains
the name of the analysed image, the number of 
colonies
detected in this image and, if a mask was used, the surface of the mask.
In the detailed output, each row of data corresponds to a different colony. 
Each colony is characterised by the name of the image it comes from,
the surface of the mask used for this image,
the position (X,Y) of its centre, its corrected median values of colour intensity,
its area, its perimeter and the number of colonies that were in the same cluster 
as
this colony. This latter output is helpful for users needing
to perform advanced analysis.

\section*{Results}

\subsection*{Speed}
\begin{figure}[!ht]

\begin{center}
\includegraphics[width=3.27in]{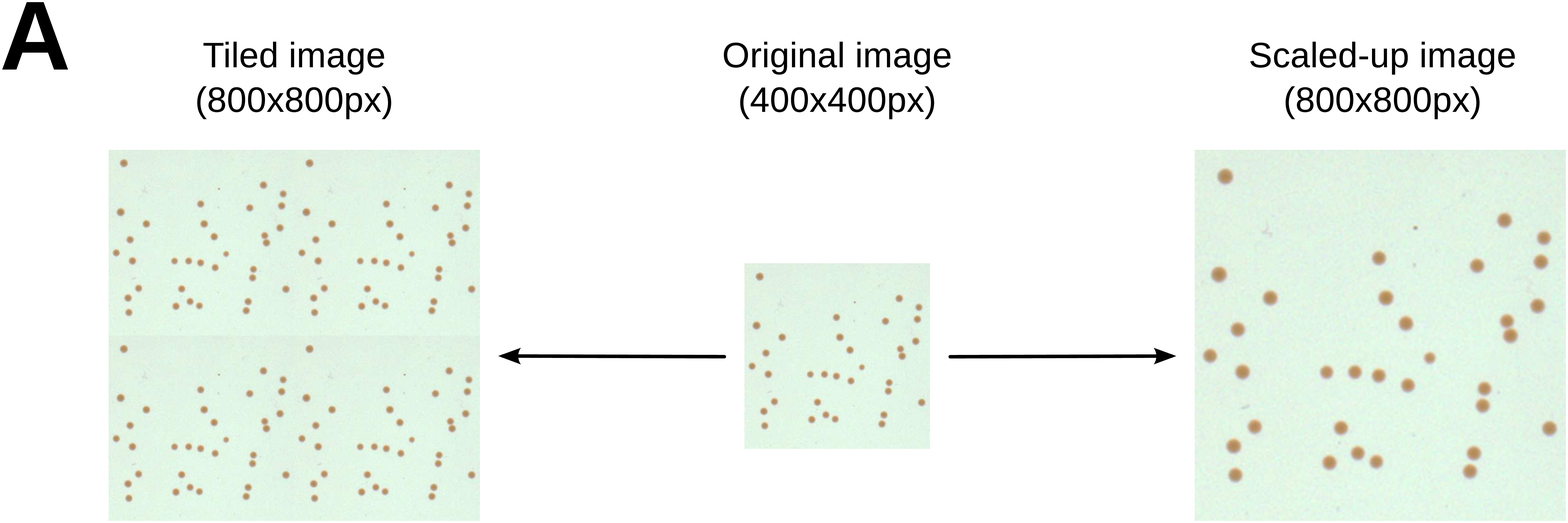}\\
\end{center}
\begin{center}
\includegraphics[width=3.27in]{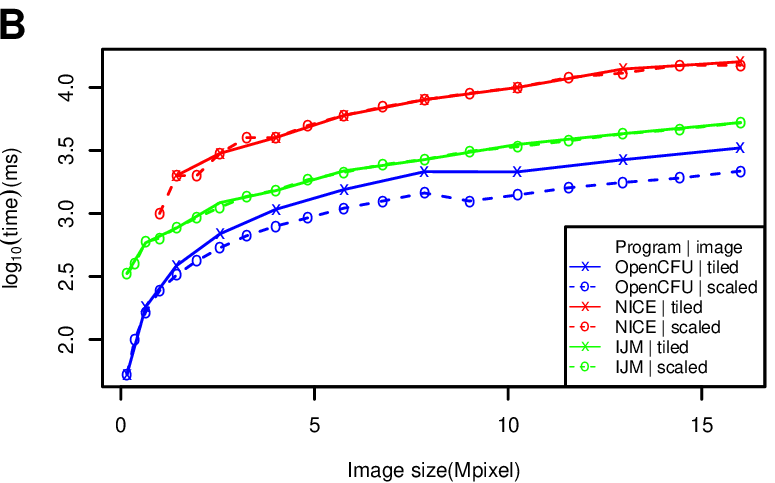}
\end{center}
\caption{{\bf Processing time of OpenCFU, NICE\cite{clarke_lowcost_2010} and an ImageJ macro\cite{cai_optimized_2011} for images of
different size.} An original arbitrary square image was either tiled to itself
or scaled-up~(A) in order to obtain a range of square images featuring an
increasing number of colonies or increasingly large colonies, respectively. The
processing time of the three methods for these images was estimated in both cases (B).
On the tested range of resolutions, OpenCFU was faster than both NICE and the ImageJ macro (\IJM). 
The segments joining points do not represent data, but only aid readability.}
\label{figSpeed}
\end{figure}

Since algorithms are likely to iteratively process each of the foreground
objects, their speed could differ according to their size and number.
Therefore, in order to assess the method speed,  two types of images were
generated from the same template:
``scaled-up'' images featuring a constant number of increasingly big bacterial
colonies, or ``tiled'' images with an increasing number of constant-sized
colonies~(fig.~\ref{figSpeed}A).

The processing time for OpenCFU, NICE\cite{clarke_lowcost_2010} and an ImageJ macro by Cai and colleagues
\cite{cai_optimized_2011}~(\IJM{}),
were compared while analysing the same images~(fig.~\ref{figSpeed}B). Under the
tested conditions, the three algorithms performed in linear time ($O(n)$).
OpenCFU was the fastest, followed by \IJM{} and NICE.
For instance, for a typical picture of 1.6$\times{}$1.6kpx the tools would take
approximatively 0.69, 1.22 and 3.0 seconds, respectively.

\subsection*{Accuracy} 
In order to assess the accuracy of OpenCFU, 19 plates
containing between 10 and 1000 \sa{} colonies were prepared, independently
enumerated by \n{} trained humans and then photographed with a high-definition camera.
The pictures were analysed by OpenCFU, NICE and \IJM{}.
The results obtained by humans and automatic methods were then compared.
The deviations of the results of each
agent from the medians of human counts, the reference, were
calculated.

NICE and, to a lesser extent, \IJM{} tended to overestimate the number of
colonies when few are present
whilst underestimating the high-density
plates 
~(fig.~\ref{figAccu}A). The significance of the slopes of the linear
regressions was assessed by a t-test:
$a = -34.56\%$~($P$-$value = 4.52 \cdot 10^{-5}$) and $a = -10.15\%$~($P$-$value =
1.98  \cdot 10^{-4}$), respectively. Analysis by OpenCFU did not result
in a significant bias: $a = -2.58\%$~($P$-$value = 0.169$).

The average of absolute deviations from the reference was used
as a measure of error and compared between agents~(fig.~\ref{figAccu}B).
Images of the same plates were also taken with a low-cost webcam and analysed in
order to estimate the impact of poor quality images on the accuracy of the three
methods.
The median error was $0.93\%$ for human agents. In order to assess the
inaccuracy of the automatic methods, their errors were compared to the human
errors by performing a Wilcoxon test:
NICE~(median error $= 9.93\%$) and \IJM{}~(median error $= 6.64\%$) had a significantly higher inaccuracy than
humans ($P$-$value = 4.65\cdot 10^{-10}$ and $1.37\cdot 10^{-5}$, respectively).
In comparison, OpenCFU~(median error $= 1.93\%$) was not less accurate ($P$-$value = 0.44$).
As expected, using poor quality pictures increased the error for OpenCFU~(median error $= 2.78\%$; $P$-$value =
0.0495$), NICE~(median error $= 13.0\%$; $P$-$value = 1.92 \cdot 10^{-10}$) and \IJM{}~(median error $= 11.1\%$; $P$-$value = 2.91 \cdot 10^{-6} $);

\subsection*{Robustness} 
In order to assess how robust the three methods were,
pictures of plates featuring typical artefacts were analysed.
Figure~\ref{figRobust}A shows qualitative results of this approach.
Both NICE and \IJM{} seemed likely to falsely count bubbles, edges, cracks or
dust, whilst OpenCFU appeared unaffected.

To quantify the robustness of the three methods to the presence of edge, the
images used for the accuracy test (fig.~\ref{figAccu}) were translated by 25 pixels to the top-left
corner to simulate a slight~(1.7mm) mispositioning during
acquisition~(fig.~\ref{figRobust}B), and were re-analysed with the same
parameters.
The bias induced was assessed by measuring the difference between the results
before and after perturbation. A paired t-test was performed between the
original results of the three methods and their respective results after perturbations.
In these conditions, \IJM{} and NICE overestimated their own result by
\EdgeNoiseIJM{} and \EdgeNoiseNICE{} colonies, respectively. OpenCFU was not affected:~\EdgeNoiseOCFU{} colonies.

In order to quantify the extent to which the presence of bubbles in the agar matrix would impact on
the precision of the methods, 18 plates containing exclusively bubbles (between
0 and 20) were analysed. The relationships between number of bubbles and number
of detected colonies is represented in figure~\ref{figRobust}C. Linear
regressions were performed and the significance of the slopes was assessed by a
t-test. The number of objects detected by \IJM{} and NICE were positively
related to the number of bubbles, \BubbleNoiseIJM and \BubbleNoiseNICE, respectively.
This was not the case for OpenCFU, \BubbleNoiseOCFU.
In order to qualitatively assess the ability of OpenCFU to process images of diverse nature,
representative pictures of circular biological objects were analysed. 
Graphical outputs are shown 
in figure~6.
These results suggest
that the algorithm is versatile enough to 
enumerate circular objects from very different images.

\section*{Discussion}
In the present study, a new algorithm based on recursive research of circular
regions over values of threshold of a grey-scale image has been presented, implemented and compared to two alternative methods\cite{clarke_lowcost_2010,cai_optimized_2011}. 
Comparison with additional software tools was considered, but, for different reasons, could not be formally performed.
Among them, Clono-counter \cite{niyazi_counting_2007}
can certainly be useful in some situations; however, it seems inappropriate for analysing numerous or large images.  
The program does not allow the user to analyse successively several files, and, for instance, using
the same mask for different images is not possible.
In addition, large images need to be down-scaled before analysis; 
as an objective comparison between programs requires all methods to process the same images,
it was not possible to use Clono-counter.
Both CHiTA\cite{bewes_automated_2008} and Arraycount\cite{kachouie_arraycount_2009} 
have probably been very useful to some researchers in the past, but
they were both unavailable for download and could not be compared to the present
method.
The method based on Cell profiler \cite {vokes_using_2008} could, in my case, not be adapted through minor
alterations in the proposed pipeline.
The ImageJ plug-in developed by Sieuwerts \emph{et al.} \cite{sieuwerts_simple_2008}
involved several human
interventions (converting to grey-scale and thresholding).
In addition, a general, rather than adaptive, threshold on the grey-scale image was, in my case, 
not satisfying since background intensity varied between and within plates.
For this reason, this plug-in was not included either.
Finally, the results presented by S. Brugger and co-workers \cite{brugger_automated_2012}
are promising, but they proposed a colony counter relying on a specific
acquisition platform rather than a standalone program capable of processing a range of 
pictures from different devices.

Despite the intensive nature of the image processing, the implementation of the described algorithm has
been shown to perform even faster than methods that are supposedly less intensive
\cite{clarke_lowcost_2010,cai_optimized_2011} (fig.~\ref{figSpeed}). This is most
likely due to the use of the optimised OpenCV library\cite{opencv_library} in
combination with custom \texttt{C++} functions.
For a standard picture, the three tested methods performed in less than three
seconds. In comparison, much slower methods, such as the one presented by Vokes
and Carpenter\cite{vokes_using_2008}, would take approximately two minutes for
the same image (data not shown).
Since, on modern hardware, the processing time will rarely be much longer than
the total acquisition time, OpenCFU's faster performance is unlikely to provide
a decisive direct gain of time over NICE\cite{clarke_lowcost_2010} and
the ImageJ macro published by Cai
and colleagues (\IJM{})\cite{cai_optimized_2011}.
However, authors have acknowledged that calibration is the most time-consuming
step of semi-automatic methods \cite{vokes_using_2008,cai_optimized_2011}.
Since, during calibration, a human will essentially try multiple values of
parameters on different sample images, a fast processing speed coupled with an
immediate display of the result will certainly provide an advantage. In
addition, OpenCFU will save calibration time by only reprocessing an image from
the first step involving the parameter that has been changed rather than from
the beginning.

\begin{figure}[!ht]
\begin{center}
\includegraphics[width=3.27in]{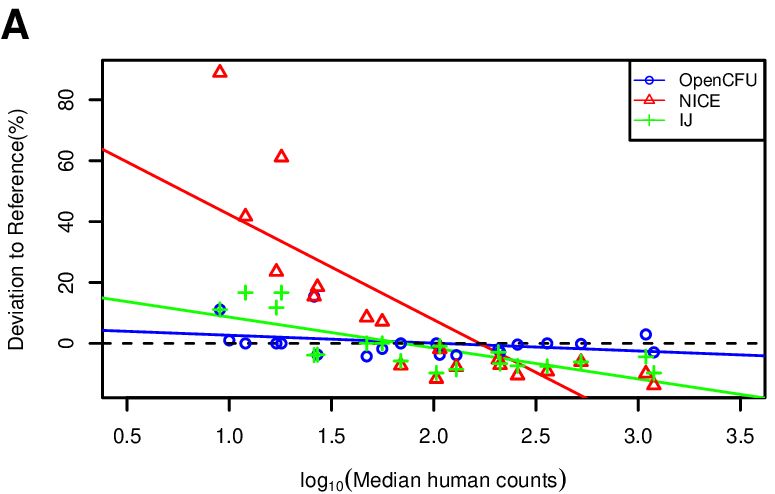}

\includegraphics[width=3.27in]{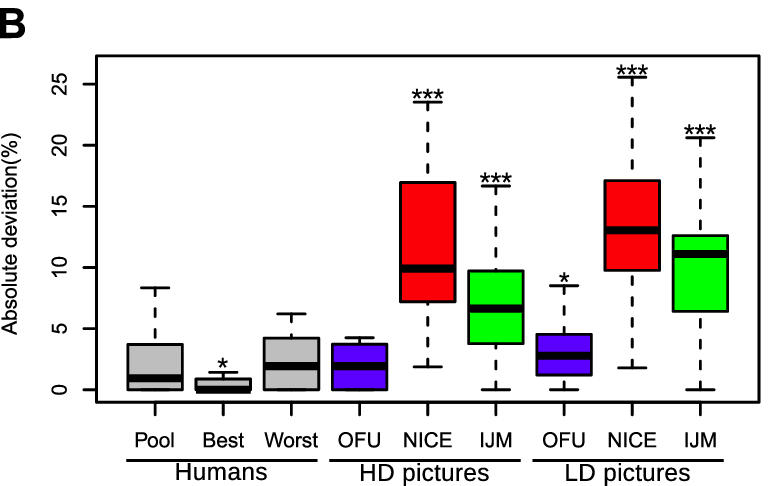}

\end{center}

\caption{{\bf Comparison of accuracy between OpenCFU, NICE\cite{clarke_lowcost_2010} and  an ImageJ macro\cite{cai_optimized_2011}.} 
The medians of \n{} humans counts were used as a reference to measure deviation. The effect of
the number of colonies on the deviation from the reference was assessed (A). For NICE and the ImageJ macro (\IJM), the
slope was significantly negative. The dotted line represents the reference.
The absolute deviation from the reference was used as a measure of error (B). Error
for the best human, the worst human and the three methods were compared to the
pooled human group. With high-definition images (HD), NICE and \IJM{} had a higher error than the pooled human
group (Pool) while OpenCFU (OFU) did not. Using low-definition pictures (LD) from a low-cost webcam increased the error for the three methods.}
\label{figAccu}
\end{figure}

\begin{figure}[!ht]
\begin{center}
\includegraphics[width=3.1in]{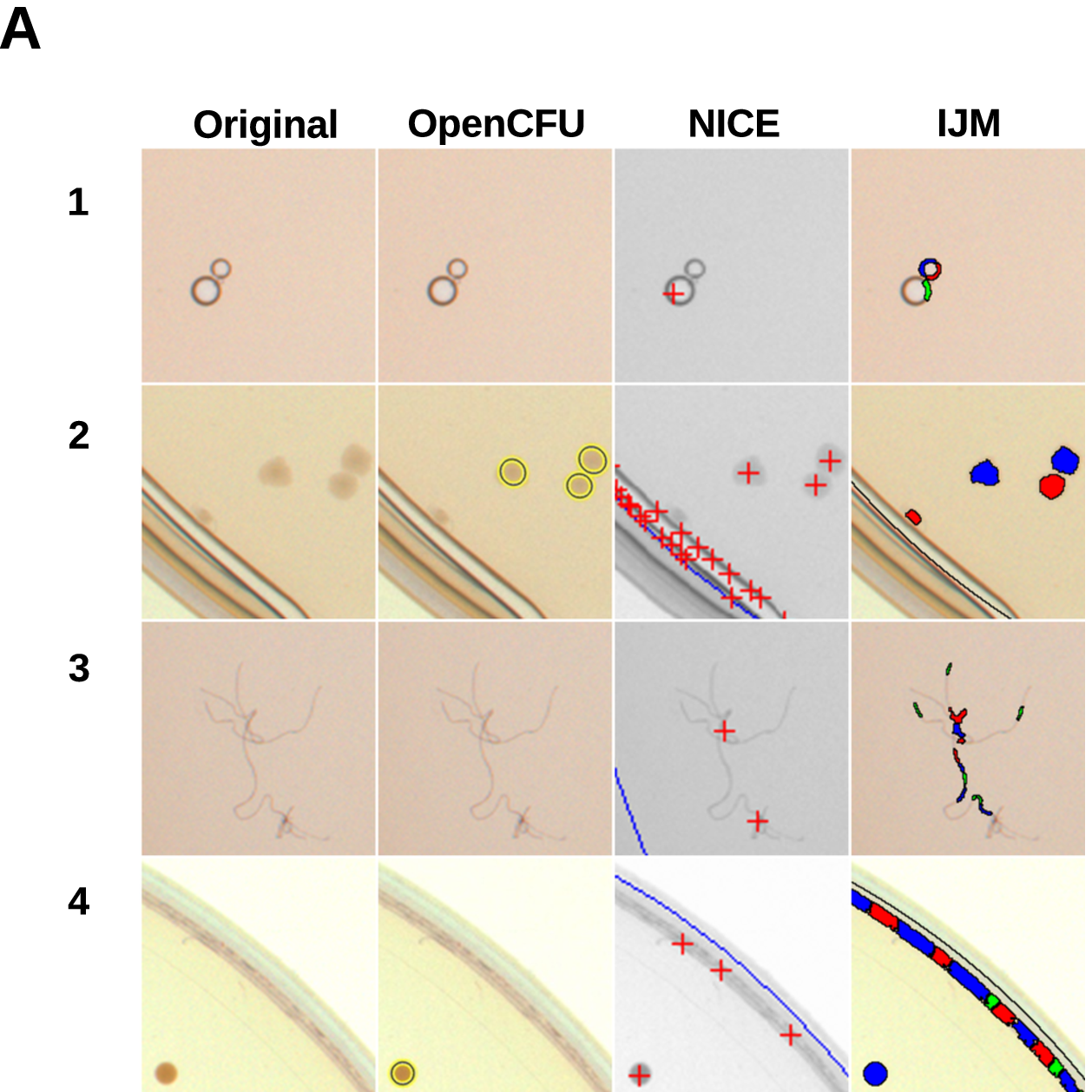}
\includegraphics[width=3.1in]{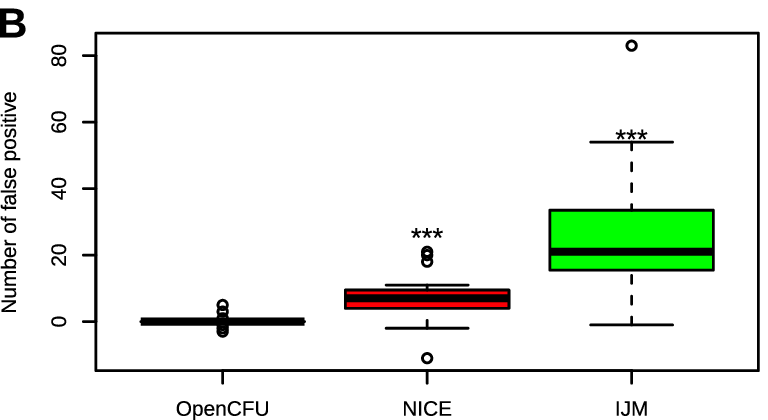}
\includegraphics[width=3.1in]{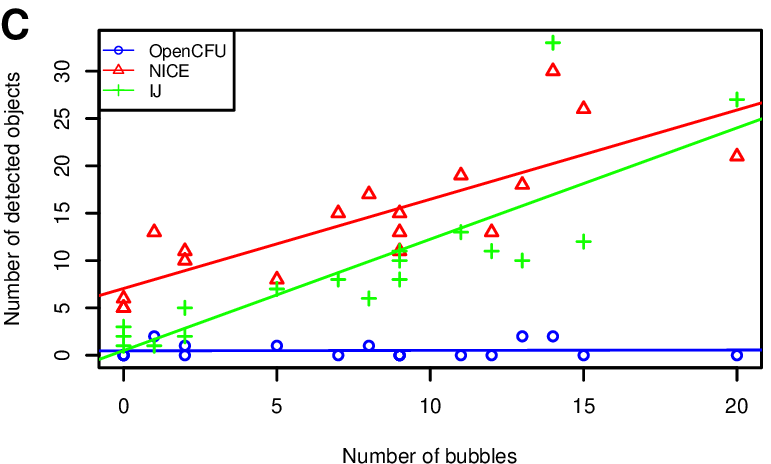}
\end{center}

\caption{{\bf Comparison of robustness to common perturbations between OpenCFU, 
NICE\cite{clarke_lowcost_2010} and an ImageJ macro\cite{cai_optimized_2011}.} 
A qualitative assessment of robustness was undertaken by analysing pictures
containing artefacts (A).
Representative portions of 1.7cm by 1.7cm (200$\times{}$200 pixels) illustrate
the results of the presence of bubbles (1), cracks in the agar (2), dust (3) and
edge of dish (4) in the region of interest. Objects detected by OpenCFU, NICE
and the ImageJ macro (\IJM) are represented by ellipses, crosses and arbitrary
colours, respectively.
NICE and \IJM{} but not OpenCFU seemed to consistently detect artefacts as
colonies.\\
A quantitative analysis of robustness to plate mispositioning was
conducted ~(B).
OpenCFU, NICE and \IJM{} were used to count the number of colonies in the
pictures of 19 plates. Then, all the images were translated by 1.7mm (25px) and
analysed with the same region of interest as the original.
This procedure induced a significant bias for NICE, \EdgeNoiseNICE{} colonies and \IJM{}
\EdgeNoiseIJM{} colonies, but not for OpenCFU \EdgeNoiseOCFU{} colonies (one-sided paired
t-test).\\
The impact of the presence of bubbles in the agar was measured by analysing
pictures of 18 plates containing exclusively bubbles~(C).
A linear regression between the number of bubbles and the number of detected
objects was performed.
NICE and \IJM{} counts were both positively related to the number of bubbles,
\BubbleNoiseNICE{} and \BubbleNoiseIJM{}, respectively. OpenCFU was not affected:
\BubbleNoiseOCFU{}	.
}
\label{figRobust}
\end{figure}

OpenCFU was shown to be very accurate and, with high-definition
pictures, did not generate more errors than the average human error~(fig.~\ref{figAccu}). In contrast, NICE and \IJM{} 
were shown to
overestimate plates with few (less than 50) colonies and underestimate plates
with large numbers 
of colonies~(fig.~\ref{figAccu}A). They also had a significantly
higher inaccuracy than humans~(fig.~\ref{figAccu}B). NICE and \IJM{} had to be
used with regions of interest drawn inside the dish. This practice excludes
colonies from the analysis and could result in a
consistent underestimation. But such bias could also be a consequence of an inefficiency to detect or segregate
 small colonies.
The overestimation of low-load plates was however probably the result of false positives
arising from the presence of artefacts which are pragmatically hard to avoid.
Using NICE with Otsu's method for thresholding gave overall good results~(fig.~\ref{figAccu}A).
However, it assumes foreground objects exist and would be inappropriate
if only few (or no) colonies are present. One
solution could be to manually count plates that have few colonies and use an
image processing-based method for high-load plates. This will surely result in
logistic complications and errors. Another solution is to individually verify
each processed image and reprocess some with a different threshold.
This is not easy with NICE because the user has to specifically click on image
names one-by-one and then click each time to query a display of the colonies.
Finally, if an image is judged to be misprocessed, there is no obvious manner to
reprocess it exclusively.
\IJM{}
does not provide a way to load a list of images and
overlay results on top of the original images for visual verification. Achieving
this is possible but involves some knowledge of ImageJ macro language.

Authors have emphasised the need to have a low-cost platform to count colonies
and have, for instance, used
desktop-scanners\cite{marotz_effective_2001,putman_simplified_2005,clarke_lowcost_2010,cai_optimized_2011
} that are generally rather slow. In this study, a low-cost webcam used in
combination with OpenCFU provided very satisfying results~(fig.~\ref{figAccu}B). OpenCFU integrates
video devices such as webcams, USB microscopes or firewire cameras in order to
facilitate this approach.

For most biologists, a moderate average deviation (lower than 20\%) will often be
negligible compared to noise generated by other experimental factors.
Rather than trying to provide a perfect similarity to human counts, methods
based on digital image processing should focus on robustness and safety.
In this study, OpenCFU has been shown to be very robust to the presence of
artefacts in comparison to alternative methods~(fig.~\ref{figRobust}A). A
simulation of a small, but likely, physical perturbation of the settings ---
mispositioning the dish during acquisition --- was proven to lead to a large
overestimation of the number of colonies by alternative
methods~(fig.~\ref{figRobust}B).
Even if a method is very accurate in perfect conditions, a biologist may be
reluctant to use it if moving the acquisition set-up by less than two
millimetres could add more than twenty colonies to the result.
In order to limit the impact of such an eventuality, the experimenter will need to
systematically check and sometimes reprocess images.
Vokes and Carpenter \cite{vokes_using_2008} solved this problem by using a
template mask and computing its best alignment to the actual image.
As OpenCFU is, by design, very robust to the presence of edges, the use of a mask is 
rarely needed. In fact, the regions of interest defined in this study
completely included the outside boundaries of the dishes.
In addition, the software provides automatic detection of a Petri dish
in
the images which corrects for shifts of the dish. This feature also avoids 
the
subjectivity of manually drawing a mask and could indirectly save acquisition
time since the user can afford to take less care in positioning the dishes.

Unlike alternatives, OpenCFU was not affected by common artefacts such as bubbles
 in the agar matrix~(fig.~\ref{figRobust}C).  
 \sa{} colonies, used in this study, are opaque and therefore produced well
contrasted areas. As NICE and \IJM{} do not perform a stringent
morphological analysis of detected objects, their sensitivity to artefacts can be
suspected to be even higher when analysing less contrasted objects.
Although 
it is possible to verify that no artefacts are present and to redraw the
mask if necessary, this step would be an additional subjective and
time-consuming human intervention.
Another advantage resulting from the robustness of the algorithm is the ability of OpenCFU
to process a diverse range of images.
Preliminary results~(fig.~6)
indicate that the algorithm is capable of enumerating different
types of colonies, as well as objects such as round seeds and pollen
from very different pictures.

In this study, humans took \humantime{} minutes in order to count the 19 plates.
Theoretically, the three methods could achieve the same result in less
than five minutes (including acquisition time). However, the lack of robustness of
NICE and \IJM{} could require systematic verification and
rectification of the results after processing. The user interface of NICE does not make this task easy and \IJM{} does not natively
provide this function. Such flaws could negate the time-gain they
would offer in perfect conditions.

OpenCFU provides postprocessing filters to eliminate detected
objects according to their intensity and colour similarity. This makes the method
even more robust since it can be used to exclude circular artefacts such as
contaminant bacteria. Most frequently, biologists will write in a small area of
the dish or in the edge; these filters can also be used to remove all writing, as it often has a very specific colour.
OpenCFU can also produce detailed output containing information about each detected
colony. This is useful for users who, for instance, wish to calculate the
respective number of distinct populations of cells.

In this study, a new algorithm was used to count bacterial colonies and
implemented in a modern and functional interface. In the tested conditions,
OpenCFU has been shown to be faster, more accurate, and
more robust to common perturbations than the two tested alternatives.
OpenCFU will help to save time and reduce subjectivity of colony
counting. Since many other biological objects (for instance, seeds, pollen, cells,
nuclei and eggs) are circular and well defined from the background, they 
can
also be counted without any modification of the software.

\section*{Availability and Future Directions}
OpenCFU is an open-source program distributed under the conditions of the GNU
General Public Licence version 3. It is available on Sourceforge (\website{}). A
user manual and video tutorial are also provided. In addition, an increasingly large list
of image samples and corresponding results will be maintained in order 
to 
help users calibrate the method. 
The program will be improved
and updated as users request features and bugs are pointed out.
Planned improvements of the program include support for multiple regions of interest,
user-supervised exclusion of outlier colonies and the availability of 
a command line version.

\section*{Materials and Methods} 

\subsection*{Image Processing} 
Details of the processing pipeline and its implementation can be freely viewed,
modified and redistributed from the source code. The file
``robustCircularRegions.cpp'' implements the new algorithm. The particle filter
and custom watershed function are in ``metaContourFilter.cpp''.
All the processing functions used are either directly provided by OpenCV or
written in \texttt{C++} (using OpenCV framework) in the same file.

The particle filter first rejects objects that do not fulfil all of the following conditions:\\
\newline{}
$Perimeter > 2\times{}minRad\times{}\pi $\\
$Area > \pi\times{}{minRad}^2$\\
$Height > minRad\times{}2$\\
\newline{}
Where,\\ 
$minRad$ is the user-defined minimal radius.

Then, it decides if a region is a single object (colony) or multiple clustered
objects. To be a single object, it must fulfil at least one of the following
criteria:\\
\newline{}
$Solidity > 0.95$\\
$Area <  1.6 \times{}\pi\times{}{minRad}^2$\\
$Perimeter < 1.6 \times{} minRad\times{}2$\\
\newline{}
Where,\\ $Solidity = \frac{Area_{shape}}{Area_{ConvexHull}}$\\
\newline{}
Otherwise, it is decided to be multiple objects.\\

If it is a single object, it must fulfil all of the following criteria:\\
\newline{}
$Aspect$ $Ratio < 1.7$\\
$Area <  \pi\times{}{maxRad}^2$\\
$\frac{Perimeter^2}{Area \times{} 4\pi } < 1.6$
\newline{}\\
Where,\\
$maxRad$ is the user-defined (or calculated from the image dimensions) maximal radius for an object.\\

If it is a cluster of objects, it must fulfil all of the following criteria:\\
\newline{}
$Aspect$ $Ratio < 5$\\
$Width > 1.4\times{}2\times{}minRad $\\
$Area < 10 \times{}\pi\times{}{maxRad}^2$\\
$\frac{Perimeter^2}{Area \times{} 4\pi } < 4.0$
\newline{}\\
To segregate clustered objects --- during the second pass of the processing ---
a distance transform is computed on a 5$\times{}$5 pixel neighbourhood with a=1, b=1.4, c=2.1969. 
The peaks of the distance transform serve as markers for a custom watershed
function. Briefly, the watershed-like function works as follows:\\
All markers correspond to the local maxima in the distance-map.\\
As long as marked regions can grow:\\  
If a marked pixel has a higher or equal value than a neighbour and the neighbour
is not marked, the neighbour becomes marked with the same label.\\
In addition, marked regions are not allowed to expand their area over a limit value $V$:\\
 $V = 1.6\times{}\pi\times{}{r}^2$ \\
 And the distance between the original marker of a region and any pixel of this region must be lower than $D$:\\
 $D = 1.6\times{}{r}$\\
 \newline{}
 Where,\\
 $r$ is the value of the corresponding peak in the distance-map (\emph{i.e.} the
 presumptive radius).

\subsection*{Plates Preparation and Manual Counting}
LB broth supplemented with 1.5\% agar was poured into 19 standard 90mm plates.
An overnight culture of \sa{} was diluted and 100$\mu{}$L were plated. The bacterial
solution was spread using ten 2mm glass beads. The plated bacteria were grown
overnight at 37$^{\circ}$.
\N{} trained individuals were given the 19 plates in a random order. The
experiment was blinded so that no subject could know the results of any other before counting.
The total time they took (excluding copying data to an electronic file) was
recorded.
Plates with bubbles~(fig.~\ref{figRobust}C) were obtained by injecting between 0.5 and 1$\mu{}$L of air in the agar matrix.
Sometimes, bubbles were surrounded by one or two smaller adjacent bubbles. Under this scenario, only the largest was counted.

\subsection*{Semi-Automatic Methods} 

OpenCFU version 3.3 was used with a ``threshold'' value of 12 for high-definition images and 7 for webcam images.
The mask was drawn automatically for all plates by choosing the option ``Auto-Petri'' with a margin of -25px. 
The software NICE and the necessary proprietary MATLAB Complier Runtime were both downloaded from
http://www.nist.gov/pml/div682/grp01/nice.cfm.
An elliptic region of interest was manually drawn and applied for all plates
in order to exclude the edge of the Petri dishes and the Otsu thresholding method
was used with the ``high resolution'' setting for high-definition pictures and ``medium resolution'' for the webcam images.
The ImageJ macro was adapted from Cai's publication\cite{cai_optimized_2011} with minor modifications.
The threshold ($z1$) was 205 and the minimal size ($z2$) was 15.

\subsection*{Hardware} 
The high quality pictures were taken using QIMAGING Micropublisher 3.3RTV
device with a TAMRON 1:1.4 25mm \o 30.5 lens. The pictures generated were
1536$\times{}$1538px well-contrasted images.
The webcam used as a capture device for the real-time enumeration was a Sweex
Blackberry Black WC250 (1600$\times{}$1200px, 30fps).
A white trans-illuminator was used to optimise contrast in both cases.

The processing time of OpenCFU, NICE and \IJM{} were assessed on the same
machine: a ``System76 Gazelle Professional'' (CPU = intel i7-2630QM,
2.00GHz).
OpenCFU and \IJM{} were used under GNU/Linux operating sytem (Linux Mint 13, kernel
3.2).
OpenCFU was compiled with \texttt{g++} 4.6.2 and OpenCV 2.4.2.
\IJM{} was used with ImageJ 1.46 (Java 1.6.0\_24).
NICE was used under Windows7-professional (64bit).

\subsection*{Statistical analysis} 
In order to assess the effect of the
number of colonies on the deviation from the reference (fig.~\ref{figAccu}A), a
linear model between deviation and $log_{10} (Reference)$, was fitted.  A t-test
was performed on the slope of the regression line.

The deviation in the count of each plate was given by:\\
\newline{}
$Deviation_p =100\times{}\frac{ (Agent_p-Reference_p)}{Reference_p}$\\
\newline{}
Where,\\
$p$ is the plate\\
and $Reference$ the median of human counts for $p$\\

The absolute deviations from the reference~(fig.~\ref{figAccu}B) were compared
between each agent and the pooled human group by performing a Wilcoxon test.
In order to assess the significance of the greater number of detected colonies
after translation of images (fig.~\ref{figRobust}B), a one-tailed paired t-test
comparing the numbers before and after perturbation was performed.
In order to quantify the effect of the number of bubbles on the number of
detected colonies~(fig.~\ref{figRobust}C), a linear model was fitted and a t-test
was performed on the slope of the regression line.
Statistical analysis was performed using \texttt{R} software\cite{R} .
 
\section*{Acknowledgments}
I am very thankful to Jens Rolff for his support and to Clayton
Costa for proving the pollen picture of 
figure~6.

\bibliography{OpenCFU_Bib}
\newpage{}

\section*{Figure Legends} 

\newpage{}

\newpage{}

\newpage{}

\newpage{}

\newpage{}

\begin{figure}[!ht]

\begin{center}
\includegraphics[width=2.27in]{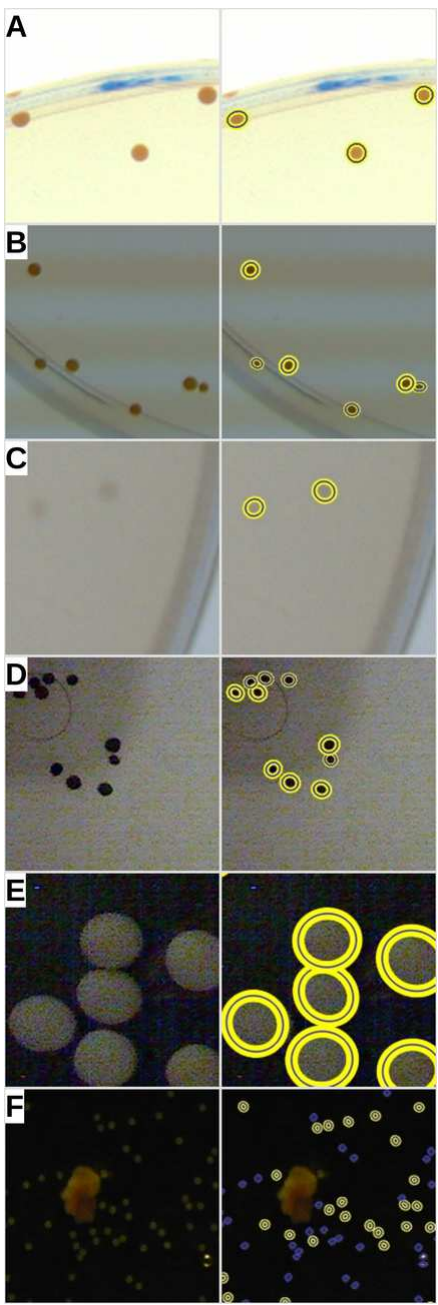}
\end{center}

\caption{{\bf Versatility of OpenCFU}.
A qualitative assessment of the versatility of OpenCFU was undertaken by 
analysing pictures of different circular biological objects:
a clear (A) and a poor quality (B) picture of \sa{} colonies,
a low-contrasted picture of \emph{Escherichia coli} (C),
a noisy picture of mustard seeds (D),
a noisy picture of soy-bean seeds (E),
and a micrography of \emph{Carduus sp.} pollen (F).
For the purpose of explanation, only representative areas (200$\times{}$200 pixels)
of six processed images are shown here.
Original portions of images are on the left and correspond to 
the graphical results obtained using OpenCFU on the right.
}
\label{figVersat}
\end{figure}

\end{document}